\documentclass[aps,prb,twocolumn,longbibliography,superscriptaddress,floatfix,10pt]{revtex4-2}
\usepackage[english]{babel}
\usepackage{amsmath,amssymb,bbm,mathrsfs,bm,braket,color,graphicx,comment,amsfonts,dsfont}
\usepackage[colorlinks,linkcolor=blue,citecolor=blue,urlcolor=blue]{hyperref}
\usepackage[mathscr]{euscript}
\usepackage{physics}
\usepackage{xcolor}
\usepackage[normalem]{ulem}
\usepackage{bm}
\usepackage{orcidlink}
\usepackage{multirow}
\usepackage{microtype}
\usepackage{bbold}
\usepackage{pifont}

\newcommand{\bs}[1]{\boldsymbol{#1}}

\begin{document}

%\title{Effective field theory for Dirac magnons transport driven by gauge fields}
\title{Transport of Dirac magnons driven by gauge fields}
\author{Luis Fernández} \email{luis.fernandez@ufrontera.cl} \affiliation{Departamento de Ciencias F\'isicas, Universidad de La Frontera, Casilla 54-D, Temuco, Chile}
\author{Ka Shen}
\affiliation{The Center for Advanced Quantum Studies and School of Physics and Astronomy, Beijing Normal University, Beijing 100875, China}\affiliation{Key Laboratory of Multiscale Spin Physics, Ministry of Education, Beijing Normal University, Beijing 100875, China}  
\author{Leandro O. Nascimento}
\affiliation{Faculdade de Física, Universidade Federal do Pará, 66075-110 Belém, Pará, Brazil}\affiliation{Universidade Federal de Campina Grande, Rua Aprígio Veloso 882, 58429-900 Campina Grande, Paraíba, Brazil}
\author{Van Sérgio Alves}
\affiliation{Faculdade de Física, Universidade Federal do Pará, 66075-110 Belém, Pará, Brazil}
\author{Roberto E. Troncoso}
\affiliation{Departamento de Física, Facultad de Ciencias, Universidad de Tarapacá, Casilla 7-D, Arica, Chile}
\author{Nicolas Vidal-Silva} \email{nicolas.vidal@ufrontera.cl}\affiliation{Departamento de Ciencias F\'isicas, Universidad de La Frontera, Casilla 54-D, Temuco, Chile}

\date{\today}% It is always \today, today,

\begin{abstract}
{\color{black}We present a unified quantum field theory for Dirac magnons coupled to emergent gauge fields. At zero temperature, any space- and time-dependent gauge perturbation drives magnons out of equilibrium, generating spin currents and magnon accumulation without conventional thermal or chemical-potential gradients. For a honeycomb ferromagnet, we derive closed-form expressions for the induced density and current. In the dc limit, the transverse spin conductivity approaches $\sigma^{xy}=\alpha^{2}\,\mathrm{sgn}(m)\hbar/4\pi$, revealing a response governed by a universal topological coefficient of the gapped Dirac spectrum and a coupling-dependent magnitude. In the ac regime, the conductivity exhibits a sharp resonance when the drive frequency matches the topological gap $\Delta$, signaling interband transitions. Our work establishes a unified gauge-invariant framework for controlling magnon transport and highlights topological signatures in gauge-driven magnon responses.}
\end{abstract}

\maketitle

\section{Introduction} 
Dirac matter refers to systems where low-energy quasiparticle excitations exhibit linear energy-momentum dispersion near Dirac points \cite{Wehling2014,Banerjee2020}. Materials such as graphene \cite{CastroRMP2009} and Dirac semimetals \cite{Wehling2014} host massless or massive Dirac fermions with relativistic-like dynamics, giving rise to phenomena including high carrier mobility \cite{Fujioka2019}, the chiral anomaly \cite{CortijoPRL2015,BurkovPRL2016,RylandsPRL2021}, and topologically nontrivial states \cite{HaldaneRMP2017,QiRMP2011}. In this context, external perturbations are captured by emergent gauge fields, introduced either via minimal coupling in the Dirac Hamiltonian or through parameter-space deformations encoded in the Berry curvature \cite{Wehling2014,Banerjee2020,VozmedianoPR2010}. These gauge fields can represent electromagnetic fields \cite{Yoshikawa2025,LiangPRL2021}, strain-induced pseudofields \cite{VozmedianoPR2010}, or spin textures \cite{Nagaosa2012,OikePRB202,LlewellynPRB2025,Tatara2019,Araki2019}, facilitating the exploration of quantum anomalies, topological responses, and transport phenomena in Dirac and Weyl systems.

Dirac magnons, the bosonic analogs of Dirac fermions, are quasiparticle excitations in two-dimensional graphene-like magnetic systems \cite{PershogubaPRX2018,Zhuo2023} and other materials \cite{ScheiePRL2022,Chung2025,Schneeloch2022}. They appear in various lattice geometries—such as honeycomb \cite{SunPRB2023,PershogubaPRX2018,Chen2021,Rintaro2025} or kagome structures—where the magnon (spin-wave) dispersion exhibits linear crossings at Dirac points. %\textcolor{red}{When time-reversal symmetry is broken, either by external magnetic fields or by the emergence of noncollinear magnetic textures, which are stabilized by Dzyaloshinskii-Moriya interactions originating from broken inversion symmetry [agregar referencias 2018 vozmediano o Cortijo 2016]}
When time-reversal symmetry is broken—either by external magnetic fields or by the emergence of noncollinear magnetic textures—Dirac magnons can acquire nontrivial topological properties, leading to intriguing phenomena such as the magnon thermal Hall effect \cite{ferreiros2018elastic,Zhuo2023}.

Currents of magnons—the net flow of spin angular momentum—are central in magnonics, due to the potential low-dissipative properties. These are generated by driving the system out of equilibrium—typically via a thermal gradients \cite{Li2025} or spin bias \cite{Brataas2020}. Emergent gauge fields offer an alternative mechanism for exciting magnon currents \cite{Tatara2019}. A prominent example is the topological magnon thermal Hall effect, where a nontrivial magnetic texture gives rise to an emergent gauge field that acts analogously to a Lorentz force on magnons \cite{Onose2010,Li2025}. This results in a transverse magnon current under a longitudinal temperature gradient. Similarly, topological magnons—arising when a magnon band acquires a non-zero Berry curvature due to time-reversal symmetry breaking and the opening of a topological gap—experience an effective pseudo-magnetic field \cite{Zhuo2023}. This field modifies magnon dynamics and transport, leading to a range of phenomena governed by topological and geometric properties of the magnon bands. In addition,
electromagnetic fields via the Aharonov-Casher effect is another type of effect over magnons mediated by gauge fields \cite{NakataPRB2019,WangPRB2024,BoliasovaPRB2025}. Weak lattice strain gives rise to elastic gauge fields give rise to pseudo Landau Levels \cite{ferreiros2018elastic,letelier2025magnons}.  While these examples showcase the role of specific gauge fields, a unified theoretical framework describing the response of Dirac magnons to general space- and time-dependent gauge drives is still lacking. Such a framework would reveal universal transport signatures independent of the gauge field's microscopic origin.

{\color{black}In this work, we propose a unified quantum field-theoretical framework for Dirac magnons coupled to emergent gauge fields. Our goal is not to revisit separately known mechanisms such as strain-induced pseudofields, Berry-curvature effects, or Hall-like magnon transport, but rather to derive within a single gauge-invariant effective-field-theory formulation the general density and current responses of Dirac magnons to generic space- and time-dependent gauge drives, independently of their microscopic origin. Within this framework, any gauge perturbation can induce magnon spin currents and accumulation, and the static and optical regimes are treated on the same footing. We further show that the induced magnon density is directly linked to both fictitious electric and magnetic fields generated by the gauge potential. Our results unveil new mechanisms for manipulating magnon transport and spin dynamics beyond conventional thermal or chemical-potential gradients. The paper is organized as follows. In Sec. II, we present the quantum field theory for Dirac magnons and derive the gauge-field-induced magnon spin current and magnon density. Section III provides a proof of concept by applying our formalism to magnons on a honeycomb lattice with next-nearest-neighbor Dzyaloshinskii–Moriya interaction, where we explore both the static and optical limits. Finally, in Sec. IV, we summarize our results and discuss the main limitations of our model.}\\
\begin{figure}
\includegraphics[width=0.4\linewidth]{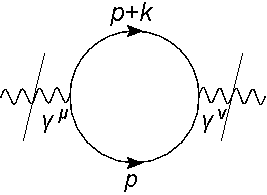}
\caption{The polarization tensor, which corresponds to the first perturbative correction of the emergent gauge field $\mathcal{A}_\mu$.}
\label{fig:diagram}
\end{figure}

\section{Model} We consider a system of Dirac magnons on a two-dimensional spin system defined in a honeycomb lattice. To capture the low-energy dynamics of Dirac magnons subject to a generic emergent gauge field $\mathcal{A}_{\mu}$, we write the minimal-coupling Lagrangian density in Euclidean space \cite{cayssol2013introduction, PhysRevX.5.011040},
\begin{equation}
\mathcal{L} = \bar{\psi} \left(\dot{\imath}\hbar\tilde{\gamma}^\mu\partial_\mu - m\right)\psi - g \bar{\psi} \tilde{\gamma}^\mu  \psi \mathcal{A}_\mu,
\label{eq:Lint}
\end{equation}
where $\psi = (\psi_A,\psi_B)^T$ corresponds to the field of magnons at the sublattice $A$ and $B$. The magnon mass, $m$, is linked to an energy gap, $\Delta$, around Dirac points, which may be related to the presence of topological states. The Dirac matrices are $\tilde{\gamma}^\mu = (\gamma^0, v_{\scalebox{0.5}{M}} \gamma^i)$, with $\mu$ running over the time ($\mu=0$) and spatial coordinates ($\mu=i=1,2$), and $v_M$ is the magnon velocity
\footnote{The magnon field is represented by $\bar{\psi}=\psi^{\dagger} \gamma^0$, where the Dirac matrices satisfy the algebra $\{\gamma^\mu,\gamma^\nu\} = -2 \delta^{\mu \nu}$, with $\gamma^0 = \dot{\imath}\sigma_z$}. The second term stands for the interaction between an emergent gauge field $\mathcal{A}_{\mu}$ and the magnonic degrees of freedom, with $g$ the coupling constant \footnote{Note that, in the chosen units, it is satisfied that $[g]=[\hbar]$, $[\mathcal{A}_0]=[\text{s}^{-1}]$, and $[\mathcal{A}_i]=[\text{m}^{-1}]$}.

The coupling to the time- and position-dependent emergent gauge field, $\mathcal{A}_{\mu}(t,\bm{r})$, drives the magnonic system out of equilibrium. We now turn to the evaluation of transport quantities that characterize the induced magnon density and magnon spin current in the linear-response regime. To start with, we consider the functional integral ${\cal Z}[\mathcal{A}]= \mathcal{N} \int \mathcal{D}\bar{\psi}\,\mathcal{D}\psi\,
\exp\left\{-\int d^3 x \mathcal{L}[\psi,\bar{\psi},{\cal A}]/\hbar\right\}$, where $\mathcal{N}$ is a normalization constant, and integrate out the magnonic degrees of freedom to obtain an effective field theory for the gauge fields.  The result can be perturbatively expanded as ${\cal Z}[\mathcal{A}] =\exp\left\lbrace S_{\text{eff}}^{(n)} /\hbar\right\rbrace$, with
\begin{align}
    S_{\text{eff}}^{(n)}=- \textrm{Tr}\sum_{n=1}\!\frac{\hbar}{n}\!\left[g\left(\dot{\imath}\hbar\tilde{\gamma}^\mu \partial_\mu\!-\!m\right)^{-1}\tilde{\gamma}^\nu \mathcal{A}_\nu\right]^n.
    \label{eq:effSn}
\end{align}
The first leading correction in the gauge field, $n=2$, corresponds to the effective action \footnote{Odd terms do not contribute to the effective action and the next term is of order $\left(g/\hbar\right)^4$, which can be neglected.},
\begin{equation}
S^{(2)}_{\text{eff}}[\mathcal{A}] =\int d^3 x \, d^3 y \, \mathcal{A}_\mu(x) {\Pi^{\mu\nu}(x\!-\!y)} \mathcal{A}_\nu(y), 
\label{eq:seff2}
\end{equation}
with the generalized coordinates $x=(t,{\bs r})$, and where we identify the polarization tensor, which in Fourier space (frequency-momentum) reads, \begin{equation}
\Pi^{\mu \nu}(p) = - \left(\frac{g}{\hbar}\right)^2\!\!\int \!\! \frac{d^3 k}{(2\pi)^3} \textrm{Tr}\left[\tilde{\gamma}^{\mu}G_0(p+k)\tilde{\gamma}^{\nu}G_0(k)\right],
\label{eq:poltensor}
\end{equation}
where $p=(p_0,\mathbf{p})$ and the free magnon propagator in the momentum space is given by
$G_0(p) = \left[\hbar\tilde{\gamma}^\mu p_{\mu} -m\right]^{-1}$. As we will describe below, the polarization tensor will determine the transport coefficient. Diagrammatically, Eq. \eqref{eq:poltensor} corresponds to the one-loop bubble shown in Fig. \ref{fig:diagram}, where the wavy lines represent the gauge field propagator.\\

The out of equilibrium response of the magnonic system under ${\cal A}_{\mu}$, traduces in an induced magnon accumulation (magnon density) and the subsequent magnon spin current. The magnon four-current $j^\mu=(\rho,j^{i})$, corresponding to the magnon density, $\rho$, and magnon current $j^{i}$, can be readily obtained from the functional differentiation of the effective action, i.e., $j^\mu(x) = \delta S^{(2)}_{\mathrm{eff}}[\mathcal{A}]/\delta \mathcal{A}_\mu(x)$ \cite{altland2010condensed}, which in Fourier space obeys,
\begin{equation}
    j^\mu(p) = \Pi^{\mu\nu}(p)\, \mathcal{A}_\nu(p),
    \label{eq:magnoncurrentgeneral}
\end{equation}
with $\Pi^{\mu\nu}(p)$ characterizing the linear response of magnons. Our results verify that $\partial_{\mu}j^{\mu}=0$, which is the expected behavior in the absence of source terms \cite{maekawa2017spin}. In terms of fictitious electric and magnetic fields
\begin{align}
\mathcal{E}_i(\omega, {\bf p}) =  \dot{\imath} \omega A_i(\omega, {\bf p}) - i p_i A_0(\omega, {\bf p})\\
\mathcal{B}_z(\omega, {\bf p}) = \dot{\imath} \epsilon^{ij}p_i A_j(\omega, {\bf p})\\
\mathcal{B}_x=\mathcal{B}_y=0,
\end{align}
 the induced magnon density reads
\begin{align}
\rho(\omega,\mathbf{p})=&\left(\frac{g}{\hbar}\right)^2\frac{1}{4\pi}\left[\frac{a(\omega,\mathbf{p})(\mathbf{p}\cdot\boldsymbol{\mathcal{E}})}{v_M^2\mathbf{p}^2-\omega^2}+2ib(\omega,\mathbf{p})\mathcal{B}_z\right],
\label{eq:magnondensity}
\end{align}
where the functions $a(\omega, {\bf p})$ and $b(\omega, {\bf p})$ are detailed in the appendix. The first contribution stems from the longitudinal application of the fictitious electric field, while the second originates from a out-of-the plane fictitious magnetic field. The former is typically associated with time-dependent external drives \cite{VozmedianoPR2010,hadadi2023pseudo,sela2020quantum}, whereas the latter is linked to the quantized magnon Landau levels \cite{li2020magnon,nayga2019magnon,sun2021magnon}. {The magnon spin current, $j^i$, can not be directly written in a closed form in terms of the fictitious field, so the general form is given in the Appendix}. However, the relevant static and optical limits, $\omega\rightarrow 0$ and $\mathbf{p}\rightarrow 0$, respectively, can be explored. For the induced static magnon current, we find
\begin{align}
\begin{split}
j^i(0,\mathbf{p})=\left(\frac{g}{\hbar}\right)^2\frac{1}{4\pi}&\Bigg[\left(\frac{p^ip^j}{p^2}-\delta^{ij}\right)a(\mathbf{p})\mathcal{A}_j(\mathbf{p})\\
&\qquad\qquad\qquad-2b(\mathbf{p})\epsilon^{ij}\mathcal{E}_j(\mathbf{p})\Bigg],
\end{split}
\label{eq:j_omega0}
\end{align}
where the fictitious electric field depends only on $A_0$. For the case $\mathbf{p}\rightarrow 0$, the optical magnon current reads
\begin{align}
\begin{split}
    j^i(\omega,{\bf0})=&\left(\frac{g}{\hbar}\right)^2\frac{i}{4\pi}\left[\frac{a(\omega)}{\omega}\delta^{ij}+2ib(\omega)\epsilon^{ij}\right]\mathcal{E}_j(\omega),
\end{split}
\label{eq:j_p0}
\end{align}
where, once the identification $j^i(\omega,{\bf0})=\sigma^{ij}(\omega)\mathcal{E}_j(\omega)$ is made, the optical magnon spin conductivity $\sigma^{ij}(\omega)$
follows directly. Interestingly, the magnon current exhibits longitudinal and Hall-like contributions—proportional to the Levi-Civita tensor—where the transverse component is induced by the pseudo–electric field.

The induced magnon density and currents, Eqs. \eqref{eq:magnondensity}–\eqref{eq:j_p0}, constitute the central results of this work. {\color{black}Their significance is that both density and current responses, as well as the static and optical regimes, are obtained from the same gauge-invariant effective kernel after integrating out the magnon degrees of freedom. In this sense, the formalism provides a unified description of Dirac-magnon transport under generic emergent gauge drives, without relying on a separate microscopic linear-response construction for each specific realization of the gauge field. We emphasize that no specific microscopic model for the magnon system has been assumed. Consequently, our theory is fully general and applies to any magnonic platform with excitations expanded around Dirac points, provided that this excitation allows for a perturbative treatment. As discussed above, the emergent gauge field may originate from strain, light, magnetic textures, mechanical rotations, and other mechanisms.} The appearance of a fictitious magnetic field in the induced magnon density captures, for instance, the strain-induced response of magnonic systems, leading to magnon Landau levels \cite{nayga2019magnon,sun2021magnon,ferreiros2018elastic,letelier2025magnons} or to Berry-curvature–driven transport in topological magnets \cite{li2020magnon,roldan2016topological,weber2022topological,owerre2016topological}. Likewise, a fictitious electric field may, in principle, emerge from time-dependent strain or external electric driving \cite{sela2020quantum,amasay2021transport,hadadi2023pseudo,sano2024acoustomagnonic}.

\section{Dirac magnons in a honeycomb lattice} To illustrate the power of our general formalism, we now apply it to ferromagnetic Dirac magnons on a honeycomb lattice. We consider a spin system aligned to the $z-$direction, and described by the spin Hamiltonian, 
\begin{align}
H_S &=-J\sum_{\langle i,j\rangle}  \mathbf{S}_i \cdot \mathbf{S}_j + \sum_{\langle\langle i,j \rangle\rangle} \mathbf{D}_{ij} \cdot \left(\mathbf{S}_i \times \mathbf{S}_j \right)-B\sum_iS^z_i. 
\label{eq:spinH}
\end{align}
It consists of an exchange coupling, $J>0$, between nearest-neighbor spins at positions $i$ and $j$, an applied magnetic field $B$ along $z$-axis, and a next-nearest neighboring Dzyaloshinskii-Moriya interaction (DMI), with DM vector $\mathbf{D}_{ij}=D_{ij}\boldsymbol{\hat{z}}$ and $D_{ij}=D\nu_{ij}$ the coupling strength. The quantized low-energy spin fluctuations, magnons, are captured through the Holstein-Primakoff transformation \cite{holstein1940field}: $S^x_i+iS^{y}_i=\sqrt{S_i-a_i^{\dagger} a_i}\,\,a_i$, $S^x_i-iS^{y}_i=a_i^{\dagger}\sqrt{S_i-a_i^{\dagger} a_i}$, and $S^z_i=S-a_i^{\dagger}a_i$, where the operator $a_i$ ($a_i^{\dagger}$) annihilates (creates) a magnon excitation at site $i$. The spin Hamiltonian $H_S$, is approximated up to quadratic on magnon operators, which in momentum space reads $H_m =  \sum_{\mathbf{p}} \psi_{\mathbf{p}}^\dagger \,\left(\Omega+\gamma_{\mathbf{p}} \cdot \boldsymbol{\sigma}\right) \,\,\psi_{\mathbf{p}}$, with $\Omega=3JS+B$ and the field operator $\psi_{\mathbf{p}} = \left( a_{\mathbf{p}}, b_{\mathbf{p}} \right)^{T}$, being $a_{\mathbf{p}}$ and $b_{\mathbf{p}}$ the annihilation operators in the sublattice $A$ and $B$, respectively. The Pauli matrices are denoted by $\boldsymbol{\sigma}$ and $\gamma_p = \sum_i \left(-JS \cos {\bf p} \cdot  \boldsymbol{\alpha}_i , JS \sin {\bf p} \cdot \boldsymbol{\alpha}_i , 2SD \sin {\bf p} \cdot \boldsymbol{\beta}_i\right)^T$, with $\boldsymbol{\alpha}$ and $\boldsymbol{\beta}$ the nearest and second neighbors of the honeycomb lattice \footnote{The nearest and second neighbors of the honeycomb lattice given by $\boldsymbol{\alpha}_{1}=a_0\left(\sqrt{3}/2,-1/2\right)$, $\boldsymbol{\alpha}_{2}=a_0\left(0,1\right)$, and $\boldsymbol{\alpha}_{3}=-a_0\left(\sqrt{3}/2,1/2\right)$; while $\boldsymbol{\beta}_{1}=-\boldsymbol{\beta}_{4}=\boldsymbol{\alpha}_{1}-\boldsymbol{\alpha}_{3}$, $\boldsymbol{\beta}_{2}=-\boldsymbol{\beta}_{5}=\boldsymbol{\alpha}_{2}-\boldsymbol{\alpha}_{1}$, and $\boldsymbol{\beta}_{3}=-\boldsymbol{\beta}_{6}=\boldsymbol{\alpha}_{3}-\boldsymbol{\alpha}_{2}$}. In the continuum limit and around the Dirac points, we get the Dirac Hamiltonian,
\begin{align}
H_{D} = \psi^{\dagger}\left( \hbar v_{\scalebox{0.5}{M}}p_x \sigma_x - \hbar v_{\scalebox{0.5}{M}}p_y \sigma_y + m \sigma_z\right)\psi,
\label{eq:HD}
\end{align}
with $v_{\scalebox{0.5}{M}} = 3 JS a_0/2\hbar$ the magnon velocity, being $a_0$ the lattice parameter. The mass term $m = 3 \sqrt{3} DS$ \cite{ferreiros2018elastic} sets the topological band gap $\Delta = 2 m$ at the Dirac points, thereby enabling the exploration of clear signatures of topological phase transitions \cite{owerre2016first}. Note that a gap can also arise from trivial mechanisms, such as sublattice anisotropy \cite{wang2018anomalous,hidalgo2020magnon}, but our focus here is on the topological case driven by DMI.

In the presence of an emergent gauge field, Eq. \eqref{eq:HD} is minimally coupled via $\partial_{\mu}\rightarrow\partial_{\mu}-ig\mathcal{A}_\mu$. The resulting response of the magnon gas depends then on the specific form of $\mathcal{A}_\mu$. We now apply the presented formalism to investigate magnon transport under the influence of the emergent gauge field in the optical and static regimes. For numerical estimates, we adopt typical parameters for two-dimensional ferromagnets \cite{aguilera2020topological,vidal2024magnonic}: 
$a_0=2.46\times10^{-10}$ m, $J=1$ meV, $D = 0.1 J$, and $S=1/2$. Since the coupling constant $g$ (with units of $\hbar$) may arise from different microscopic mechanisms, we parametrize it as $g=\alpha\hbar$, where $\alpha$ is a dimensionless factor encoding the effective coupling between the gauge and magnon fields. For elastic deformations, it is proportional to the strain tensor and the Grüneisen parameter \cite{ferreiros2018elastic, nayga2019magnon}. In contrast, for magnons propagating on a spin-textured background in the presence of DMI, the gauge field acquires a form proportional to the ratio $D/J$ \cite{kim2019tunable, li2020magnon, roldan2016topological}. Rotating systems provide another example, where magnons experience fictitious forces encoded in emergent spin gauge fields proportional to temporal variations of the rotation angle \cite{funato2025spin, fujimoto2020magnon}. Finally, we assume a momentum-independent Gilbert damping parameter $\alpha_G$ \cite{BrataasPRL2008,KamraPRB2018,SimensenPRB2020}.
\begin{figure}[h]
    \centering  \includegraphics[width=1\linewidth]{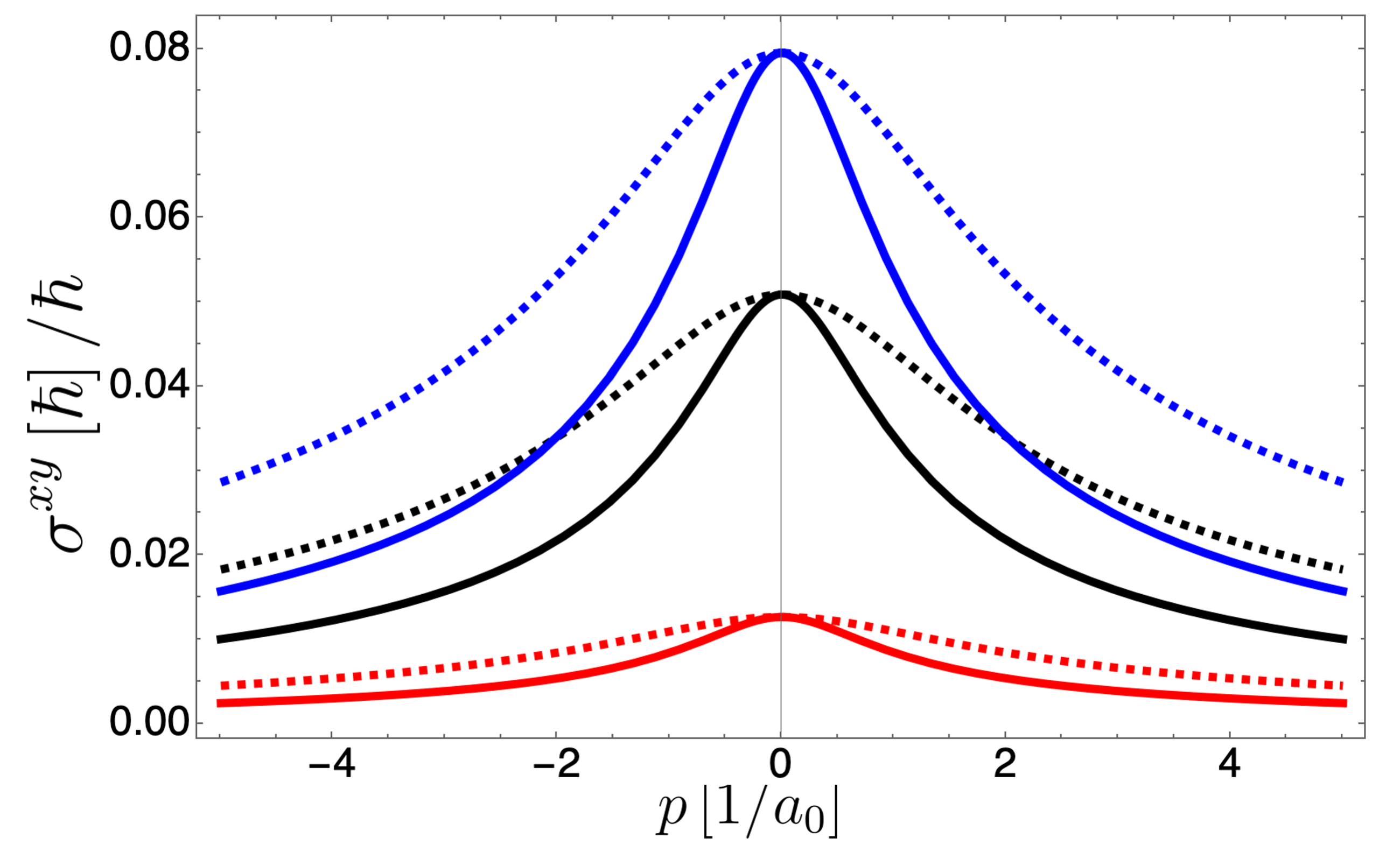}
    \caption{Momentum dependence of the transverse spin conductivity $\sigma^{xy}$ for different values of the coupling constant $\alpha$ and group velocities $v_M$. Blue, black, and red lines correspond to $\alpha=1, 0.8, \text{and }0.4$, respectively, while dashed lines stand for $v_M=v_M/2$. {\color{black}The curves approach their maximum dc value as \(p \to 0\), followed by a decay with increasing momentum. The reduction in the magnon group velocity \(v_M\) suppresses this decay, thereby producing a broader low-momentum plateau.}}
\label{fig:sigma_static}
\end{figure}

\subsection{Static limit} The static conductivity is represented by the limit $\omega\rightarrow 0$ (Eq. \eqref{eq:j_omega0}), and finite momentum. First, we consider ${A}_{0}\neq 0$ and ${A}_{x}={A}_{y}= 0$. This case is equivalent to introducing control over the magnon chemical potential \cite{demidov2017chemical}. We note that only a transverse component of the conductivity contributes to the magnon current, and is given by
\begin{equation}
\begin{split}
    \sigma^{ij}(p) = \frac{\alpha^2}{2\pi} \epsilon^{ij} \frac{m}{v_{\scalebox{0.5}{M}}\left|{\bf p}\right|} \arcsin\left[\frac{\hbar v_{\scalebox{0.5}{M}}\left|{\bf p}\right|}{\sqrt{4 m^2 + \left(\hbar v_{\scalebox{0.5}{M}}\right)^2{\bf p}^2}}\right], 
    \end{split}
\end{equation}
where $\sigma^{xy}=-\sigma^{yx}$. In Fig. \ref{fig:sigma_static}, we show $\sigma^{xy}$ for different values of the coupling constant $\alpha$ and group velocity, as a function of momentum $p = \sqrt{p_x^2+p_y^2}$. {\color{black} One can note the symmetry with respect to its maximum value at $p = 0$, for which the conductivity approaches
\begin{align}
    \sigma^{xy}(p\to 0)=\alpha^{2}\,\hbar\,\mathrm{sgn}(m)/4\pi ,
\end{align}
corresponding to the dc transverse response. Analogously to graphene \cite{CortijoPRL2015}, this expression contains a coefficient of topological origin, which in the present case is \((\hbar/4\pi)\mathrm{sgn}(m)\). However, in contrast to a strictly universal quantized conductivity, its overall magnitude is modulated here by the effective gauge coupling \(\alpha^{2}\), which depends on the microscopic origin of the emergent gauge field. Therefore, the topological structure of the response is universal, while the full conductivity is not strictly universal in magnitude. This coefficient plays a role analogous to the electronic conductance scale \(e^{2}/4\pi\hbar\) in graphene, but adapted here to the magnonic response. We emphasize that this transverse response relies on the topological nature of the Dzyaloshinskii-Moriya-induced gap; a trivial one (e.g., one arising from sublattice anisotropy) would not yield the same response structure. Also, as can be seen from Fig. \ref{fig:sigma_static} , the conductivity decreases for larger values of the momentum modulus, and such a behavior is less pronounced for smaller values of \(v_M\). It is worth noting that the transverse response derived here requires a finite Dirac mass. In the limit of a vanishing mass, the Berry curvature becomes zero and the transverse conductivity disappears, reflecting the topological origin of the effect. This suggests that edge-sensitive geometries, such as nanoribbons, may provide favorable platforms to probe the transverse response \cite{owerre2016topological}.}

Let us now consider the case $A_0=0$ and ${A}_i\neq0$, which may arise from strain-induced lattice modulations. In this regime the fictitious electric field vanishes, and the magnon current contains longitudinal and transverse components. According to Eq. (\ref{eq:j_omega0}), it takes the form $j^{i}=\eta^{ij}A_j$, where
\begin{align}
\eta^{ij}=\frac{\alpha^2}{4\pi}&\Bigg[\left(\frac{p^ip^j}{p^2}-\delta^{ij}\right)a(\mathbf{p})\Bigg].
\end{align}
In  particular, if $A_y=0$, the ficticius magnetic field is ${\cal B}_z=-ip_yA_x$, and the induced longitudinal magnon current becomes $j^x=(i\eta^{xx}/p_y){\cal B}_z$ and the transverse $j^y=(i\eta^{yx}/p_y){\cal B}_z$. %{\color{blue}Note that both longitudinal and transverse magnon current response with a different structure from the conventional notion of a fictitious electric field and, conversely, from usual spin conductivity.} 
Note that both longitudinal and transverse magnon current responses exhibit structures distinct from those associated with a fictitious electric field and, likewise, from conventional spin conductivity. Our result shows that this response persists even in the absence of magnon mass and is governed primarily by the function 
$a(\mathbf{p})$ defined above. Indeed, when 
$m=0$ we have the function $a(\mathbf{p})=p (\pi/2)v_M$, so the magnon current remains finite. {\color{black}In addition, for a constant fictitious field \(B_z\), the dc limit \((\omega \to 0,\, p \to 0)\) yields the induced magnon density
\begin{align}
\rho_{\mathrm{DC}}=(\hbar/4\pi)\alpha^{2}\,\mathrm{sgn}(m)\,B_z \, ,
\end{align}
which again reflects the topological coefficient associated with the gapped Dirac spectrum, modulated by the effective gauge coupling \cite{liu2021strain}.}

\subsection{Optical limit} Here we consider the limit $\mathbf{p}\rightarrow 0$ and $\omega\neq 0$ to obtain the AC response. From Eq. \eqref{eq:j_p0}, we get both the longitudinal and transverse magnon spin conductivities
\begin{align}
\sigma^{xx}(\omega) &\label{eq:longitudinalAC}= \frac{\dot{\imath}\alpha^2}{4\pi\omega} \left[\left(\frac{\hbar^2\omega^2 +4 m^2}{4\hbar\omega}\right) \zeta(\omega)-\left|m\right|\right]\\
\sigma^{xy}(\omega) &\label{eq:transversalAC} =  \frac{m \alpha^2}{2\pi \omega}\zeta(\omega)
\end{align}
where $\sigma^{xx}=\sigma^{yy}$ and $\sigma^{xy}=-\sigma^{yx}$, and $\zeta(\omega)=\text{arcsinh} \left[{\hbar\omega} /\sqrt{4m^2 -\hbar^2\omega^2}\right]$.
\begin{figure}[h]
    \centering  \includegraphics[width=1\linewidth]{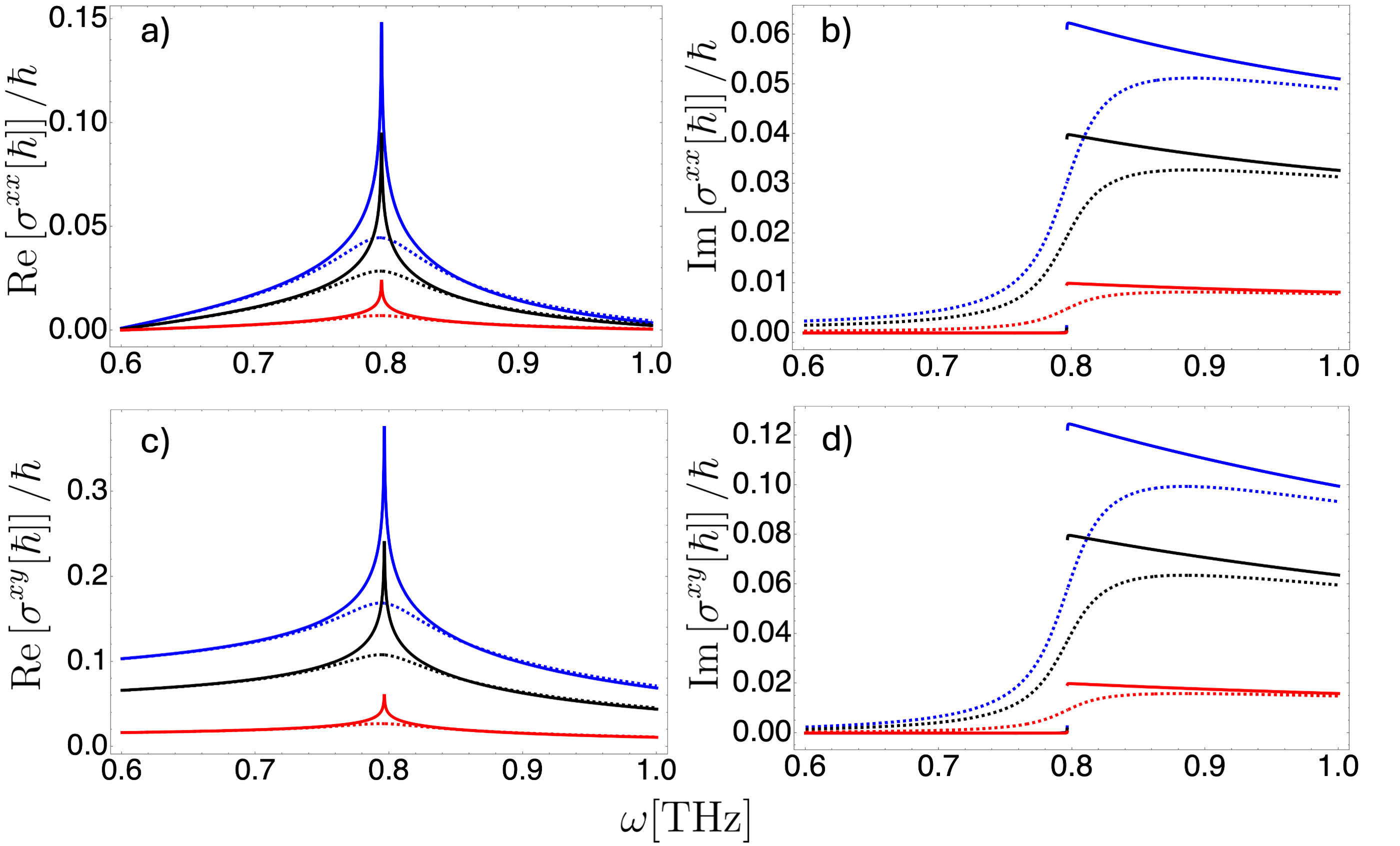}
    \caption{AC magnon spin conductivity as a function of the frequency $\omega$. Panel a) and b) show the real and imaginary part of the longitudinal response $\sigma^{xx}$, while c) and d) show the real and imaginary part of the Hall-like conductivity $\sigma^{xy}$. Blue, black, and red lines correspond to $\alpha=1, 0.8, \text{and }0.4$, respectively. Solid lines correspond to a Gilbert damping $\alpha_G = 10^{-5}$, while dashed lines stand for $\alpha_G = 3\times 10^{-2}$. Resonance occurs at the frequency determined by the magnon gap, i.e.,  $\Delta/\hbar=6\sqrt{3}DS/\hbar\approx 7.96\times 10^{2}$ GHz and the dissipation Gilbert damping $\alpha_G$ broadens this resonance. The imaginary parts reveal the threshold behavior associated with interband transitions for $\omega>\Delta/\hbar$.}
    \label{fig:sigma_AC}
\end{figure}
To incorporate dissipation and regularize the response at resonance, we introduce the phenomenological Gilbert damping $\alpha_G$ by the replacement $\omega\rightarrow \omega(1+\dot{\imath} \alpha_G)$ in Eqs. \eqref{eq:longitudinalAC} and \eqref{eq:transversalAC}, following standard treatments of magnon damping \cite{BrataasPRL2008,KamraPRB2018,SimensenPRB2020}. In Fig. \ref{fig:sigma_AC} we show the real and imaginary parts of $\sigma^{xx}$ (panels a) and b)) and $\sigma^{xy}$ (panels c) and d)) as a function of the frequency drive. Note that the real part, related to the effective spin transport, is resonantly enhanced at $\omega = \Delta/\hbar$, i.e., at the topological gap $\Delta=6\sqrt{3}DS/\approx 0.5196$ meV. Such a resonance is broadened by the inclusion of dissipation $\alpha_G$. On the other hand, for the clean case ($\alpha_G = 0$), the imaginary part is non-nulled only for $\omega>2m/\hbar$, which demonstrates that interband transitions require an energy $\hbar\omega>\Delta$. This threshold behavior can be interpreted as a magnonic analog of optical absorption in gapped Dirac materials. Indeed,  while the Hall-like conductivity $\sigma^{xy}$ vanishes for $m=0$, the longitudinal response $\sigma^{xx}$ converges to a purely imaginary constant $\sigma^{xx}=i\alpha^2\hbar/32$, which is indicative of a closing gap and the absence of interband transition. For $\alpha_G\neq 0$, the magnon band is broadened so the interband conversion can take place for values distinct from $\Delta$. The imaginary parts of the conductivity, shown in Figs. \ref{fig:sigma_AC}b) and \ref{fig:sigma_AC}d), are related to the out-of-phase (reactive) response of the magnon gas. For instance, $\Im{\sigma^{xx}}$ is connected to the magnon density of states and the system's polarization response, while $\Im{\sigma^{xy}}$ indicates a phase lag in the transverse current relative to the driving . {\color{black}At finite temperature, Bose-Einstein occupation factors are expected to reweight the magnon bands, while additional temperature-dependent linewidth effects should further smooth the interband threshold and broaden the optical resonance. Nevertheless, the characteristic frequency scale of the ac response remains set by the gap \(\Delta\), so that the resonance position continues to track the underlying gapped Dirac spectrum.  In this sense, the present zero-temperature treatment should be understood as describing the intrinsic-response regime, in which thermal occupation effects are negligible compared with the gap scale. For the bandgap considered here ($\Delta = 0.5196\,\mathrm{meV}$), the corresponding thermal scale is $\Delta/k_B \approx 6\,\mathrm{K}$. Hence, the two-band description applies in the low-temperature regime $T \ll \Delta/k_B$, consistent with neutron-scattering experiments reporting well-defined single-magnon excitations in honeycomb magnets at temperatures of order $6\,\mathrm{K}$ \cite{chung2025massive}. }

\section{Conclusions} We have developed a unified effective-field-theory framework for Dirac-magnon transport in the presence of emergent gauge fields. We find that both electric and magnetic fictitious fields generate magnon spin currents and magnon density responses. Because such gauge fields can arise from time-dependent external drives (e.g., electromagnetic fields or lattice rotations) as well as from topological features encoded in magnetic textures or mechanical deformations, our approach naturally incorporates a broad class of driving mechanisms. {\color{black} From a broader perspective, several ingredients appearing in our model, such as Berry-curvature effects, magnon Hall transport, and strain-induced gauge fields, have been discussed separately in previous works. However, the purpose of the present work is not to reproduce these results within a different language, but rather to provide a unified quantum field-theoretical framework describing the coupling between Dirac magnons and generic emergent gauge fields. Within this approach, the magnon density and spin current are obtained directly from the effective action after integrating out the magnon degrees of freedom. This formulation naturally yields the response to space- and time-dependent gauge fields in a compact and gauge-invariant form, allowing the transverse spin response and the optical resonances to be derived systematically without relying on a specific microscopic linear-response construction.  It is worth mentioning that only the longitudinal spin conductivity will have corrections at higher-order ($\sim \alpha^4$) and these
are dependent on the dynamics of the gauge-field. The transverse spin conductivity, nevertheless, has no higher-order correction \cite{marino2015interaction}}

{\color{black}We predict a pronounced optical conductivity resonance when the drive frequency matches the topological gap, \(\Delta = 2m\), reflecting an interband transition. In addition, the transverse dc response approaches $
\sigma^{xy}=\alpha^{2}(\hbar/4\pi)\,\mathrm{sgn}(m)$,
which contains a universal topological coefficient associated with the gapped Dirac spectrum, modulated by the effective gauge coupling. In this way, the topological structure of the response is universal, while its full magnitude depends on the microscopic mechanism generating the emergent gauge field. Note that our model also allows for exploring Hall and valley currents according to the specific choice of the gauge field $\boldsymbol{\mathcal{A}}$.\\

The present analysis has been carried out at zero temperature in order to isolate the intrinsic response of Dirac magnons to emergent gauge fields. At finite temperature, the predicted response is expected to be reweighted by the Bose-Einstein occupation of magnon bands, leading primarily to a reduction in magnitude and to a broadening of the ac resonant features, while the underlying mechanism associated with the gapped Dirac spectrum should remain unchanged. A full finite-temperature treatment lies beyond the scope of the present work and will be addressed in future work. Thus, our results establish a gauge-field-based route to generate and control spin currents through the topological structure of gapped Dirac magnons, beyond thermal or magnon chemical potential gradients.}

\section{Acknowledgments}  L.F. acknowledges
the financial support from Vicerrectoría de Investigación y Postgrado de la Universidad de La Frontera through Project No. BPVRIP-092024. R.E.T. and N. V-S. thank funding from ANID Fondecyt Regular 1230747 and 1250364, respectively. L. F. and N. V-S thank Prof. Javier Lorca for the fruitful discussions.  V. S. A. and L. O. N.  are partially supported by Conselho Nacional de Desenvolvimento Científico e Tecnológico-Brazil (CNPqBrazil), process: 408735/2023-6 CNPq/MCTI

\appendix

\section{Polarization Tensor}

The polarization tensor defined in Eq.~\ref{eq:poltensor} is calculated using the $2\times2$ representation of the Dirac matrices $\tilde{\gamma}^\mu=(\gamma^0,v_M\gamma^i)$, for which the trace operations are given by

\begin{align}
\mathrm{Tr}[\tilde{\gamma}^\mu \tilde{\gamma}^\nu] &= -2\delta^{\mu\nu}, \\
\mathrm{Tr}[\tilde{\gamma}^\mu \tilde{\gamma}^\alpha \tilde{\gamma}^\nu] &= 2\epsilon^{\mu\alpha\nu}, \\
\mathrm{Tr}[\tilde{\gamma}^{\mu} \tilde{\gamma}^{\alpha} \tilde{\gamma}^{\nu} \tilde{\gamma}^{\beta}]
&=
2\left(
\delta^{\mu\alpha}\delta^{\nu\beta}
-\delta^{\mu\nu}\delta^{\alpha\beta}
+\tilde{\delta}^{\mu\beta}\delta^{\nu\alpha}
\right).
\end{align}

These identities, together with the Feynman parameterization \cite{Peskin1995},

\begin{equation}
\frac{1}{a^n b^m}
=
\frac{\Gamma[n+m]}{\Gamma[n]\Gamma[m]}
\int_0^1 dx\,
x^{n-1}(1-x)^{m-1}
[b+x(a-b)]^{-(n+m)},
\end{equation}

are used with

\begin{align}
a &= (p_0+k_0)^2 + v_M^2(\mathbf{p}+\mathbf{k})^2 + m^2,\\
b &= k_0^2 + v_M^2\mathbf{k}^2 + m^2 .
\end{align}

After straightforward algebra we obtain

\begin{widetext}
\begin{equation}
\Pi^{\mu\nu}(p)=-
\left(\frac{g}{\hbar}\right)^2
\int_0^1 dx
\int\frac{d^dk}{(2\pi)^d}
\frac{\Theta^{\mu\nu}}
{\left[k_0^2-v_M^2\mathbf{k}^2+x(1-x)(p_0^2-v_M^2\mathbf{p}^2)-m^2\right]^2}.
\end{equation}
\end{widetext}

For the different choices of Lorentz indices $\mu(\nu)=0,i$, three distinct numerators $\Theta_{\mu\nu}$ arise.\\

For $\mu=\nu=0$:
\begin{align}
\Theta^{00}&= 2\left[k_0(k_0+p_0)-\delta^{ij}v_M^2 k_i(k_j+p_j)
-m^2\right].
\end{align}
For $\mu=0$, $\nu=i$:
\begin{align}
\Theta^{0i}&=2v_M^2\left\{\delta^{ij}\left[k_0(k_i+p_i)+k_i(k_0+p_0)\right]+\epsilon^{ij}mp_j\right\}.
\end{align}
For $\mu=i$, $\nu=j$:
\begin{align}
\nonumber\Theta^{ij}&=-2v_M^2\Bigl\{
\delta^{ij}k_0(k_0+p_0)-v_M^2\Big[k_i(k_j+p_j)+k_j(k_i+p_i)\\
&-\delta^{ij}\mathbf{k}\cdot(\mathbf{k}+\mathbf{p})
\Big]+\delta^{ij}m^2-
\epsilon^{ij}m\,p_0\Bigr\}.
\end{align}

Defining $p^2=p_0^2+v_M^2\mathbf{p}^2$, the polarization tensor components read

\begin{widetext}
\begin{equation}
\Pi^{00}(p)=-
\left(\frac{g}{\hbar}\right)^2
\frac{1}{4\pi}
\frac{\mathbf{p}^2}{p^2}
\left[
|m|
+
\frac{p^2-4m^2}{2\sqrt{p^2}}
\arcsin
\left(
\frac{\sqrt{p^2}}{\sqrt{4m^2+p^2}}
\right)
\right].
\end{equation}
\begin{equation}
\Pi^{0i}(p)=
\left(\frac{g}{\hbar}\right)^2
\frac{1}{4\pi}
\left\{
\frac{p^0p^i}{p^2}
\left[
|m|
+
\frac{p^2-4m^2}{4\sqrt{p^2}}
\arcsin
\left(
\frac{\sqrt{p^2}}{\sqrt{4m^2+p^2}}
\right)
\right]
-
2\epsilon^{ij}m
\frac{p_j}{\sqrt{p^2}}
\arcsin
\left(
\frac{\sqrt{p^2}}{\sqrt{4m^2+p^2}}
\right)
\right\}.
\end{equation}

\begin{equation}
\Pi^{ij}(p)=-
\left(\frac{g}{\hbar}\right)^2
\frac{1}{4\pi}
\left\{
\left(
\delta^{ij}-v_M^2\frac{p^ip^j}{p^2}
\right)
\left[
|m|
+
\frac{p^2-4m^2}{2\sqrt{p^2}}
\arcsin
\left(
\frac{\sqrt{p^2}}{\sqrt{4m^2+p^2}}
\right)
\right]
+
2\epsilon^{ij}m
\frac{p_0}{\sqrt{p^2}}
\arcsin
\left(
\frac{\sqrt{p^2}}{\sqrt{4m^2+p^2}}
\right)
\right\}.
\end{equation}
\end{widetext}

This effective model preserves gauge invariance, which satisfies

\begin{eqnarray}
p_0\Pi^{0\nu}(p)+v_M p_i\Pi^{i\nu}(p)=0.
\end{eqnarray}

This condition implies the conservation of the current and ensures that the gap opened by the Dzyaloshinskii--Moriya interaction remains stable. The magnon density and spin current are obtained from Eq. \eqref{eq:magnoncurrentgeneral} as

\begin{widetext}
\begin{equation}
\rho(\omega,\mathbf{p})=
\left(\frac{g}{\hbar}\right)^2
\frac{1}{4\pi}
\left[
\frac{i\mathbf{p}^2 a(\omega,\mathbf{p})}{\omega^2-v_M^2\mathbf{p}^2}
A_0(\omega,\mathbf{p})
+
\frac{i\omega p^i a(\omega,\mathbf{p})}{v_M^2\mathbf{p}^2-\omega^2}
A_i(\omega,\mathbf{p})
-
2\epsilon^{ij}p_j b(\omega,\mathbf{p})A_i(\omega,\mathbf{p})
\right].
\end{equation}

\begin{eqnarray}
\nonumber j^i(\omega,\mathbf{p})&=
\left(\frac{g}{\hbar}\right)^2
\frac{1}{4\pi}
\Big[
\frac{\omega p^i a(\omega,\mathbf{p})}{v_M^2\mathbf{p}^2-\omega^2}
A_0(\omega,\mathbf{p})
-
\left(
\delta^{ij}-\frac{v_M^2 p^i p^j}{v_M^2\mathbf{p}^2-\omega^2}
\right)
a(\omega,\mathbf{p})
A_j(\omega,\mathbf{p})
+
2i\epsilon^{ij}p_j b(\omega,\mathbf{p})A_0(\omega,\mathbf{p})\\
&-2i\epsilon^{ij}\omega b(\omega,\mathbf{p})A_j(\omega,\mathbf{p})
\Big].
\end{eqnarray}
\end{widetext}

with

\begin{align}
a(\omega,\mathbf{p})
&=
\sqrt{m^2}
+
\left(
\frac{v_M^2\mathbf{p}^2-\omega^2-4m^2}
{\sqrt{v_M^2\mathbf{p}^2-\omega^2}}
\right)
F(\omega,\mathbf{p}),\\
b(\omega,\mathbf{p})
&=
\frac{m}{\sqrt{v_M^2\mathbf{p}^2-\omega^2}}
F(\omega,\mathbf{p}),
\end{align}

and

\begin{equation}
F(\omega,\mathbf{p})=
\arcsin
\left[
\frac{\hbar\sqrt{v_M^2\mathbf{p}^2-\omega^2}}
{\sqrt{4m^2+\hbar^2v_M^2\mathbf{p}^2-\hbar^2\omega^2}}
\right].
\end{equation}
\bibliography{biblio}

\end{document}